\documentclass[%
 preprint,
nofootinbib,
 amsmath,amssymb,
 aps,
]{revtex4-1}

\usepackage{geometry}                        
\usepackage{graphicx}                
\usepackage{amssymb}
\usepackage{latexsym}
\usepackage{comment}
\usepackage{color}
\usepackage{amsmath}
\newcommand{\cE}{{\cal E}}
\newcommand{\ep}{\epsilon}

\newcommand{\tcr}{\textcolor{black}}
\newcommand{\Sm}{{S_m}}
\newcommand{\E}{\mbox{I}\hspace{-0.2em}\mbox{E}}

\begin{document}
\preprint{APS/123-QED}
\title{Dynamics of bacterial populations under the feast-famine cycles}
\author{Yusuke Himeoka and Namiko Mitarai}
\affiliation{The Niels Bohr Institute, University of Copenhagen, Blegdamsvej 17, Copenhagen, 2100-DK, Denmark}
 \email{yusuke.himeoka@nbi.ku.dk and mitarai@nbi.ku.dk}
\date{\today}

\begin{abstract}
Bacterial populations in natural conditions are expected to experience stochastic environmental fluctuations, and in addition, environments are affected by bacterial activities since they consume substrates and excrete various chemicals. We here study possible outcomes of population dynamics and evolution under the repeated cycle of substrate-rich conditions and starvation, called the “feast-famine cycle”,  by a simple stochastic model with the trade-off relationship between the growth rate and the growth yield or the death rate. In the model, the feast (substrate-rich) period is led by a stochastic substrate addition event, while the famine (starvation) period is evoked because bacteria use the supplied substrate.  Under the repeated feast-famine cycle, the bacterial population tends to increase the growth rate, even though that tends to decrease the total population size due to the trade-off. Analysis of the model shows that the ratio between the growth rate and the death rate becomes the effective fitness of the population. Hence, the functional form of the trade-off between the growth and death rate determines if the bacterial population eventually goes extinct as an evolutionary consequence. We then show that the increase of the added substrate in the feast period can drive the extinction faster. Overall, the model sheds light on non-trivial possible outcomes under repeated feast-famine cycles.
\end{abstract}

\maketitle
\section*{Introduction}
As already pointed out in the 18th century, exponential growth is the most prominent feature of population dynamics \cite{malthus1872essay}, and bacterial systems are probably the best-studied model system about exponential growth. However, as pointed out by  J. Monod \cite{monod1949growth}, the exponential growth is only one of the growth phases of bacteria, and the stationary phase, the death phase, and the lag phase are all as important for the bacterial population dynamics. 

A variety of theoretical studies on the bacterial population dynamics tend to focus on the competition for nutrients under constant environment \cite{gause1969struggle,pianka1970r}, where the competition takes place mainly in the form of exponential growth under a constant influx of substrate (combined with dilution/death to keep the environment constant).  While these models have provided fundamental insights into bacterial population dynamics, in natural environments like ponds, soils, and puddles, the nutrients may be supplied by rarely happening events rather than continuous influx. Under such natural environments, bacteria experience substrate rich conditions and poor conditions alternately.

This cycle between substrate rich and poor conditions is called the feast-famine cycle \cite{hengge1993survival,simon2002microbial,datta2016microbial,kellerman2014chemodiversity,seymour2017zooming,savageau1983escherichia,merritt2018frequency}. In contrast to the continuous nutrient supply (or the constant environment) condition, under the feast-famine cycle there is no steady-state for the amount of the substrate, and accordingly for the number of the cells. While the environment becomes substrate-rich for some time after the substrate addition event, once the cells in the environment run out all the substrates, they have to tolerate until next substrate addition event which is typically highly stochastic. 
The feast-famine cycle is more than just a fluctuating environment, in that the rate of the substrate consumption affects the feast and famine period. Cells starve until the substrate is supplied, and once the environment gets substrate-rich, the cells use it quickly. During the feast (substrate-rich) period the growth of the cells changes the state of the environment. If cells use the substrate slowly, the feast period lasts longer and vice versa. 

In this sense, the feast-famine cycle is one example that cells are not just affected by the environmental condition, but also changing the environment. While the population dynamics without such feedback are well studied \cite{kussell2005phenotypic,kussell2005bacterial}, our understandings with the feedback between the environment and the population are still underdeveloped. 

\tcr{
In the seminal Lenski and coworkers’ experiment of long-term bacterial evolutionary adaptation \cite{lenski1991long},  bacteria have been diluted into fresh media every 24 hours while the growth reaches the saturation after less than 10 hours, hence the cells do experience repeated feast-famine cycles. However, the famine period is rather short that there is no visible death during the period, and the major observed adaptation was to increase the growth rate and decrease lag-time during the feast period, without a significant increase of the death rate during the famine period \cite{vasi1994long}. Interestingly, however, in a separate experiment, Vasi and Lenski had isolated mutants after the $30$ to $49$ days starvation \cite{vasi1999ecological}, and most of them are found to be inferior in competing with their progenitors in fresh medium, but having better resistance to starvation. One of the five strains clearly showed an inferior fitness to its original strain for a one-day growth competition but had superior survivability in 15-day starvation. These observations suggest that evolutionary adaptation under repeated feast-famine cycles with long-enough famine periods could result in fairly non-trivial results, especially if there is a trade-off between survivability in the long famine period and the growth rate in the feast period. It is also worth mentioning that, the trade-off of growth rate and survivability has been clearly documented for {\it E. coli} mutants with genetically manipulated $rpoS$ activity \cite{yang2019temporal}; the {\it rpoS} gene encodes for the stationary/starvation sigma factor $\sigma^S$ and known to cause trade-offs in multiple stressed conditions \cite{notley2002rpoS,maharjan2013form}.  
}

\tcr{
In the previous works,  the trade-off between the growth rate and the growth yield are well-documented \cite{jasmin2012yield,pfeiffer2001cooperation,ferenci2016trade,novak2006experimental,maclean2008tragedy,kreft2004biofilms,frick2003example}. Fast-growth but low-yield results in an extended starvation time and a reduction of the population size at the end of the feast period. It has been theoretically predicted that a fast-growth but low-yield species will wipe out a slow-growth but high-yield species in a well-mixed environment \cite{pfeiffer2001cooperation,kreft2004biofilms,frick2003example}. However, the effect of the trade-off between the growth rate and the survivability has not been theoretically analyzed yet. 
}

\tcr{
In this paper, we construct a population dynamics model where the population growth is driven by a discrete and stochastic substrate addition events. In order to understand the simplest case, we consider a well-mixed system with a single niche, i.e, there is only one kind of nutrient/substrate that bacteria can consume to grow.  Bacteria cells divide by consuming substrates, and if there is no substrate the cells die at a constant rate. In addition to the growth/death dynamics, mutations take place to change the growth rate. We introduce a trade-off between the growth rate and the death rate or the growth yield:  The fast-growing cells have a higher chance to win the competition for the nutrients in the feast period, but instead, are less tolerant in the famine period because of high death rates or a small number of the population due to the low yield.
}

\tcr{
 Using the model, we show that the bacterial population faces the Tragedy of The Commons (TOC)" type phenomenon \cite{hardin1968tragedy} under feast-famine cycles; the fast-grower tends to take over during the feast period, making the population less tolerant for the famine period. This evolutionary dilemma drives the population to faster growth and eventual extinction in the famine period for the growth-yield trade-off as previously predicted \cite{pfeiffer2001cooperation}, but we find that in the growth-death rate trade-off case, the long-term outcome depends on the functional form of the trade-off. By analyzing the model, we show that this is because the effective fitness under repeated feast-famine cycle is determined by the ratio of the growth rate and the death rate.  We then focus on the non-trivial case of the growth-death rate trade-off where functional form drives the population to extinction in repeated feast-famine cycles, and show that the TOC effect can be prompted by increasing the amount of substrate added to the environment. Finally, we discuss the model assumptions and possible extensions in comparison with experimental results in the literature.  
}

\section*{Model}
The model consists of a state vector $\vec{X}$ defined by $(M+1)$ integers that denote the population of $M$ species (either genotypes or phenotypes) and the number of substrates in the environment. It is expressed as $\vec{X}=(N_0,N_1,\cdots,N_{M-1},S)$, where $N_i$ is the number of the $i$th bacteria, and $S$ represents the number of the substrates. All elements of the state vector ${\vec X}$ are non-negative integers. Each species can have a different growth rate, growth yield, and death rate. 

A single individual of the $i$th species proliferates at a constant rate $\mu_i$ being given by $\mu_i=(i+1)\Delta\mu$ if $S$ is larger than zero, 
while it dies at rate $\gamma_i$ under the starving condition. Each species has a different growth yield $Y_i$, and $\lceil 1/Y_i\rceil$ substrates are consumed when a single bacterium of the $i$th species divides where $\lceil \cdot \rceil$ is the ceiling function. The number of the substrate in the environment is recovered to $S=\Sm$ when the substrate addition event takes place at a constant rate of $1/\lambda$. 

We introduce the mutation among species to the model. It occurs with probability $\rho$ when an individual divides, and then, the daughter cell of the $i$th species becomes either the $(i-1)$th or $(i+1)$th species in an equal probability.

By assuming that each event occurs as the Poisson process, the master equation for the simplest one species case is given by
\begin{eqnarray}
\frac{dP(N,S)}{dt}&=&\mu(N-1)P(N-1,S+\Delta S) - (1-\theta(S))\mu N P(N,S)\nonumber\\
&+&\theta(S)\gamma\Bigl((N+1)P(N+1,0)-NP(N,0)\Bigr)\label{eq:single}\\
&-&P(N,S)/\lambda+\delta_{S,\Sm}\sum_{i=-\infty}^{\Sm}P(N,i)/\lambda \nonumber
\end{eqnarray}
where $P(N,S)$ is the probability of the state with $N$ bacteria and $S$ substrates. $P(N,S)=0$ holds if $N$ is smaller than zero or $S$ is larger than $\Sm$. $\delta_{i,j}$ is Kronecker's delta, and $\hat{\delta}_{i,j}$ is given by $(1-\delta_{i,j})$. 

$\theta(S)$ is unity for $S\leq 0$ else zero. $\Delta S$ is given as $\lceil 1/Y\rceil$ (the results do not change by modifying the model so that the number of the substrate to be consumed is determined stochastically with the average consumption per division as $1/Y$.). The states with negative values of $S$ are able to have non-zero values, while the right hand side of the equations for the states with negative $S$ are the same with that of $S=0$. We regard the sum of the probabilities with $S\leq 0$ as the probability of the starving state. This model construction is to avoid dividing the equation into cases that $S$ is larger and smaller than $\Delta S$ (for detailed description, see Appendix A).

Fig.\ref{fig:fig1}a shows a realization of stochastic dynamics generated by the present model Eq.(\ref{eq:single}) with the Gillespie algorithm \cite{gillespie1977exact}. The oscillation in the number of bacteria results from the feast-famine cycle.

\section*{Results}
\subsection*{The Tragedy of Commons dilemma in the bacterial evolution under the feast-famine cycle}

We study the effect of the feast-famine cycle in the multi-species system The master equation for $M\ (M>1)$ species system is obtained by just extending Eq.(\ref{eq:single}) for $M$ species and introducing mutation among the species. There is no direct interaction among the species, but the species interact via the competition for the substrate. For the exact expression, see Eq.~(\ref{eq:meq_multi}) in Appendix. We compare the dynamics of the model with the growth rate-yield ($\mu-Y$) and the growth rate-death rate ($\mu-\gamma$) trade-off separately.

\begin{figure}[htbp]
\begin{center}
\includegraphics[width = 140 mm, angle = 0]{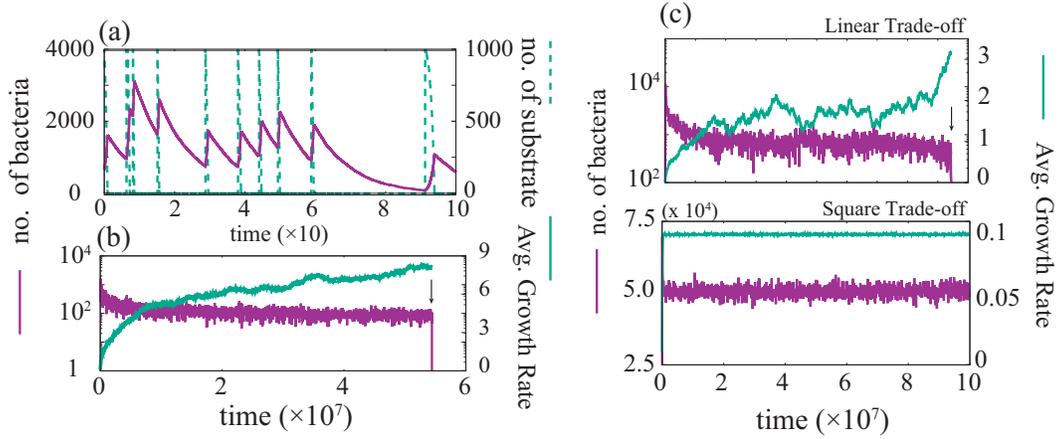}
\caption{(a). An example of the dynamics of one-species model (Eq.(\ref{eq:single})). The number of bacteria oscillates driven by the feast-famine cycle. (b). Evolution simulation with the growth-yield trade-off model. $Y=1/(\kappa+\mu)$ where $\kappa$ is introduced to avoid the divergence of the function. The population-averaged growth rate keeps increasing and the whole population extinct at the point indicated by the black arrow. (c). Evolution simulations with two different choices of the growth-death trade-off. (top) With the linear trade-off, $\gamma=a+b\mu$, the population-averaged growth rate keeps increasing and the whole population extinct at the point indicated by the black arrow. (bottom). On the other hand, the growth rate and the number of bacteria get stable at a certain value with the square trade-off, $\gamma=a+b\mu^2$. The stable growth rate is predicted as $\sqrt{b/a}=0.1$ by the analysis of Eq.(\ref{eq:Mmap}) which corresponds well with the numerical result.  While the same parameter values of $a$ and $b$ are used for the two trade-off relationships, the outcome does not change qualitatively even if different values are used for them. The parameters are set to $\mu=1,\gamma=0.101,\lambda=10,$ and $\Sm=10^3$ for (a), and $a=10^{-3},b=0.1,\delta\mu=10^{-2},\kappa=10^{-2},\rho=10^{-3},N_{\rm ini}=100,\lambda=1.28$, $\Sm=128$, $\gamma=0.15$ (only for (b)) and $Y=1$ (only for (c)) for (b) and (c). }
    \label{fig:fig1}
  \end{center}
\end{figure}

We carried out stochastic simulations of the $M$-species model (Eq.(\ref{eq:meq_multi})) with a fixed $\lambda$ and $\Sm$ value using the Gillespie algorithm. Fig.\ref{fig:fig1}b and c show evolution time courses with several trade-off relationships. There are initially $N_{\rm ini}$ cells with the lowest growth rate $\mu_0$ and the corresponding yield or the death rate, and $\Sm$ substrates. In the $\mu-Y$ trade-off case (Fig.\ref{fig:fig1}b, with trade-off $Y=1/(\kappa+\mu)$ with a constant $\kappa$),  the population-averaged growth rates keep increasing by evolution, and eventually, the whole population collapses. It is consistent with previous reports arguing that under the growth-yield trade-off, the species with a higher growth rate outcompetes others even though it leads to a reduction of the population size due to the small yield \cite{pfeiffer2001cooperation,kreft2004biofilms,frick2003example}. On the other hand, in the $\mu-\gamma$ trade-off case, the evolution of the growth rate can either leads to population collapse (Fig.\ref{fig:fig1}c top, with linear trade of $\gamma=a+b\mu$ with constant $a$ and $b$) or reach a steady state depending on the form of the trade-off (Fig.\ref{fig:fig1}c bottom, with square trade-off $\gamma=a+b\mu^2$). 

To understand the differences of the outcome, we constructed a simplified version of the model (Eq.(\ref{eq:meq_multi})) which allows us analytical calculations. We approximate that the population size is a continuous quantity and consider deterministic growth and death. We denote the number of the $i$th species right before the $n$th substrate addition event by a continuous variable $N_i(n)$. Also, we regard $S$ as a continuous variable, and thus, remove the ceiling function from the yield. 

After the $n$-th substrate addition, the species grow exponentially until all the substrate runs out. The length of the feast period after the $n$th addition event, $\tau(n)$, is determined by $\Sm=\sum_{i=0}^{M-1}N_i(n)/Y_i(\exp[\mu_i\tau(n)]-1)$, because the increment of the total population divided by the growth yield should sum up with the added substrate $\Sm$. If the interval between the $n$th and the $(n+1)$th addition events is longer than $\tau(n)$, the cells experience the famine period to die at the rate $\gamma_i$. Thus, the number of cells right before the $(n+1)$th substrate addition event is given by
\begin{equation}
        N_i(n+1)=
    \begin{cases}
    N_i(n)e^{\mu_i \tau(n)}e^{-\gamma_i(\Delta t(n)-\tau(n))}\ (\tau(n)<\Delta t(n))\\
     N_i(n)e^{\mu_i \Delta t(n)}\ ({\rm otherwise}),\\
    \end{cases} 
    \end{equation} 
where $\Delta t(n)$ is the stochastic variable representing the interval between the $n$th and the $(n+1)$th addition events, which follows the exponential distribution with average $\lambda$. By taking average of the effective growth rate, $\ln(N_i(n+1)/N_i(n))$, over the exponential distribution, we obtain a deterministic, discrete map system which describes the dynamics of the population growth as
\begin{eqnarray}
    N_i(n+1)&=&N_i(n)\exp(\hat{\mu}_i\lambda)\nonumber \\
    \hat{\mu}_i(n)&=&\mu_i\Bigl(1-\exp(-\tau(n)/\lambda)\Bigr)-\gamma_i\exp(-\tau(n)/\lambda).\label{eq:Mmap}
\end{eqnarray}
In Appendix, we show that the map dynamics has at least $M$ fixed points that only one species exists and the number of cells is zero for the other species, which is given by
\begin{equation}
N_i^{\rm st}(\mu_i)=\frac{\Sm Y_i}{(1+\gamma_i/\mu_i)^{\mu_i \lambda}-1}.\label{eq:sol_map}
\end{equation}

The linear stability analysis for the fixed points showed that only one fixed point among the $M$ fixed points is stable. The condition for the fixed point to be stable is to have the largest ratio of the growth rate to the death rate $\mu_i/\gamma_i$ (see Appendix).

In the $\mu-Y$ trade-off case, the death rate $\gamma_i$ has no index dependency, and thus, the stability or the fitness is simply determined by the growth rate. There, the growth yield $Y_i$ has no effect on the competition among the species, and thus, the bacterial population evolves to increase the growth rate without caring of how the growth is efficient in terms of the substrate consumption.

In contrast, the stability criterion tells us that the functional form of $\mu-\gamma$ trade-off affects whether the evolution lasts until the whole population goes extinct or not. When the trade-off is linear ($\gamma=a+b\mu$), the ratio $\mu/\gamma$ has no upper bound, while the ratio is bounded in the square trade-off ($\gamma=a+b\mu^2$) case. Therefore, in the square trade-off case, once the growth rate reaches the optimal point (the maximum $\mu/\gamma$), the system stays at that state. On the contrary, due to the lack of the maximum, the growth rate will never stop increasing for the linear trade-off case and leads to the population collapse. 

In both the growth-yield trade-off and the linear growth-death trade-off case, the growth rate increases over generation while the tolerance to the famine period gets worse due to the decrease or increase of the yield or the death rate. The evolution then leads to the reduction of total population size. Eventually, the population size becomes too small to tolerate fluctuations of the substrate addition time, and it results in the extinction of the whole population when the substrate addition times are longer than usual by chance.  This can be seen as one of the typical consequences of "The Tragedy of The Commons" Dilemma.

\subsection*{Impact of the feast-famine cycle for the survival of the bacterial population}
In the previous section, we have seen that under a repeated feast-famine cycle, the TOC dilemma is evoked if the trade-off relationship is such that the death rate increase linearly or slower than linear with the growth rate, or when there is a growth rate-yield trade off. In the following sections, we study how the "degree" of the feast-famine cycle affects the survival of the bacterial population and the evolutionary dynamics when the TOC dilemma is evoked. We chose to focus on the less trivial case of linear growth-death trade-off model

$\gamma(\mu)=a+b\mu$
in the following.

Firstly, we introduce the "degree" of the feast-famine cycle. In the following, we change the value of the average famine period $\lambda$ while keeping the time-averaged substrate supply ${\bar S}=\Sm/\lambda$ constant. With this constraint, the change in $\lambda$, and accordingly in $\Sm$, controls the severeness of the feast-famine cycle. A large $\lambda$ (and $\Sm$) value indicates that a large amount of the substrate is supplied to the environment less often, corresponding to the severe feast-famine cycle. On the contrary, the limit of $\lambda\rightarrow0$ with keeping $\bar S$ constant would correspond to the continuous substrate-supply limit, though strictly speaking in the present model this limit cannot be taken due to the discreteness of $\Sm$. 

\begin{figure}[htbp]
\includegraphics[width = 150 mm, angle = 0]{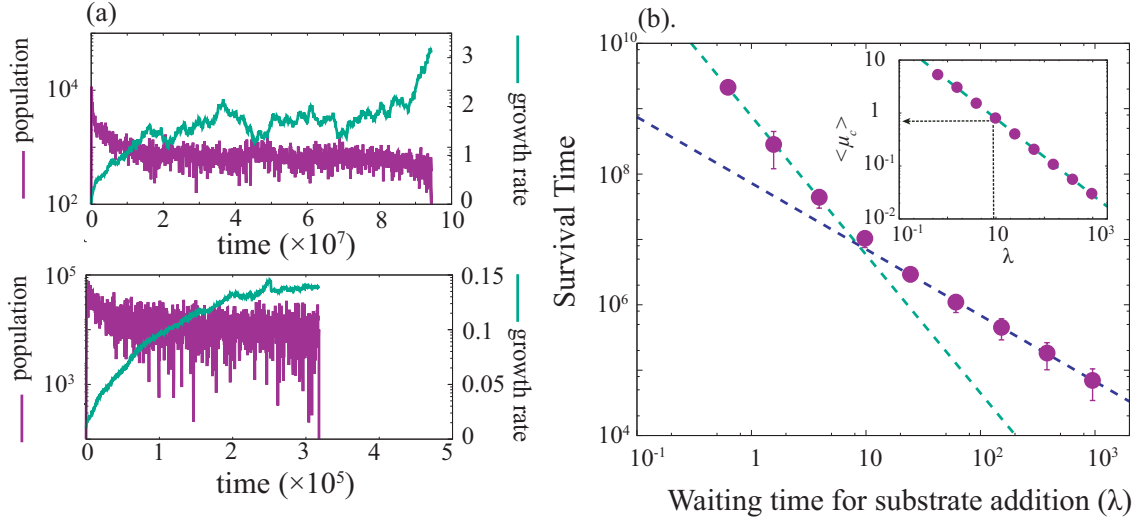}
\begin{center}
\caption{(a). The time courses of the total population and the average growth rate. (top) Under a weak feast-famine condition ($\lambda=1.28$), and (bottom) a strong feast-famine condition ($\lambda=81.92$). Extinction takes place in much shorter time under the strong feast-famine condition than the weak one. (b). The averaged extinction time is plotted with the standard deviation for several $\lambda$ values. The average and the standard deviation are computed from $128$ extinction events. $\Sm$ is given as $\Sm=\lambda\bar{S}$. The two slopes were obtained by fitting. The population-averaged growth rate achieved is also plotted in the inset. Parameter values are set to be $a=10^{-3},b=0.1,\delta\mu=10^{-2},\rho=10^{-3},$ and $\bar{S}=100$.}
    \label{fig:fig2}
  \end{center}
\end{figure}

We compared the dynamics under the different degrees of the feast-famine cycle. Fig.\ref{fig:fig2}a shows two time courses of the population and the averaged growth rate under a moderate (top panel, $\lambda=1.28$) and a severe (bottom panel, $\lambda=81.92$) feast-famine cycle. The population-averaged growth rates commonly evolve to increase over time, and eventually, the whole populations go extinct as expected from the foregoing analysis. The difference of the degree of the feast-famine cycle appears in the length of time to extinct, which we call the survival time $T_s$, and the population average growth rate just before the extinction, which we call the critical growth rate $\langle \mu_c\rangle $.

We plotted the survival time $T_s$ and the critical growth rate $\langle \mu_c\rangle $ as a function of the degree of the feast-famine cycle ($\lambda$) in Fig.\ref{fig:fig2}b. The survival time and the critical growth rate decrease as $\lambda$ increases, reflecting the harsher environment. $\langle \mu_c\rangle$ decreases approximately proportional to $1/\lambda$, while interestingly the survival time $T_s$ shows a cross-over from $T_s\propto \lambda^{-2}$ to $T_s\propto \lambda^{-1}$ at $\lambda\approx 10$.

Qualitatively, the shorter survival time $T_s$ and the smaller critical growth rate $\langle \mu_c \rangle$ with increasing 
$\lambda$ is the reflection of the asymmetry of the growth and the death in this setup. Increasing $\lambda$ increases the possible population growth per feast period linearly with $\lambda$ because of the increase of $S_m=\bar S \lambda$, but the death in the famine period affects the population exponentially as a factor $\exp(-\gamma \lambda)$. Clearly, the death effect is dominant, hence it is harder to survive with longer $\lambda$, resulting in shorter $T_s$ and smaller $\langle \mu_c \rangle$.
Quantitative analysis requires more careful consideration, which we present in the next section.\\

\subsection*{Crossover from the directed evolution to neutral evolution}
In order to have a better understanding of the observed behavior, we now focus on the cross-over of the survival time $T_s$ shown in Fig.\ref{fig:fig2}b from being approximately proportional to $\lambda^{-2}$ to $\lambda^{-1}$.

In the moderate feast-famine cycle ($\lambda \ll 10$) depicted in the top panel of Fig.\ref{fig:fig2}a, it appears that the evolution speed of the growth rate has two regimes: The evolution speed significantly slows down after the average growth rate reaches $\approx 1$, which happens around time $1\times 10^7$ in this example. Since the mutation probability is constant, we hypothesized that the dynamics of how a new species takes over the majority of the population changes with the average growth rate. 

To quantify this change of the dynamics, we studied the dynamics of taking-over among species by setting $N_{\rm ini}$ cells of the $i$th phenotype at $t=0$, and ran the population dynamics until the dominant species becomes another. 

For the simulations performed in this section, we add the spontaneous migration term to the model. The spontaneous migration makes it possible to increase the number of the cells by one without consuming any nutrient even under the starved condition (for the detailed expression of the term, see Eq.(\ref{eq:meq_multi})). An introduction of the term is a mathematical treatment for gathering many trajectories efficiently for better statistics. Since the spontaneous migration rate $\epsilon$ is set to be sufficiently smaller than any other parameters in the model, it is effectively the same as introducing one cell into the system when the population extinct. 

\begin{figure}[htbp]
\begin{center}
\includegraphics[width = 100 mm, angle = 0]{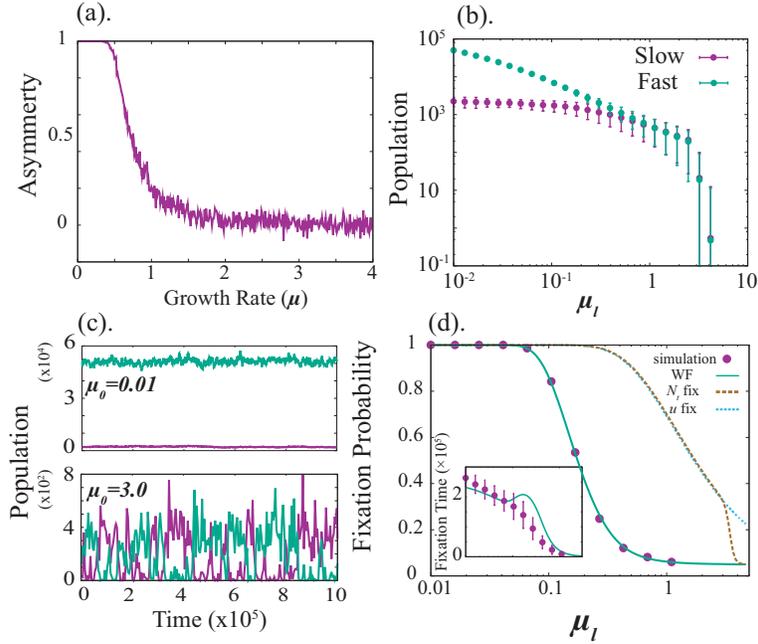}
\caption{(a). The asymmetry of the evolution. The evolution of the growth rate is directed up to $\mu\approx1$, while the growth rate behaves similarly to the random walker in a larger $\mu$ region. (b). The averaged population is plotted against the growth rate of the slow grower ($\mu_l$) with the error bar as the standard deviation. The growth rate of the fast grower is set to be $5\%$ higher than that of the slow grower. (c). Examples of time courses with $\mu_l=0.01$ (top) and $0.64$ (bottom). (d). The comparison of the fixation probabilities obtained from the numerical simulation and the calculation of the WF model. The comparison of the fixation time is also shown in the inset.  For (d), we set $\ep=\rho=0$ so that one of the two species eventually be fixed. We choose the initial value of the total number of the bacteria as $N_{\rm st}(\mu_l)$ given in Eq.(\ref{eq:sol_meq}), and the initial fraction of the species with faster growth rate as $5\%$. Parameters are set to be $\lambda=1.0,\Sm=100,a=10^{-3},b=10^{-1},\rho=10^{-3}$, and $\epsilon=10^{-8}$.}
    \label{fig:fig3}
  \end{center}
\end{figure}

From this computation, we obtained the transition probabilities from the $i$th species to another. 
With our default parameter set, it never happened that the species other than the nearest neighbors of $i$ becomes dominant before $i-1$th or $i+1$th dominates the system. Thus, the obtained transition probabilities were for increasing the growth rate by $\Delta \mu$ (probability $p(\mu)$) or decreasing it by $\Delta\mu$ (probability $1-p(\mu)$). Fig.\ref{fig:fig3}a shows the asymmetry of the probability for increasing/decreasing the growth rate, 
defined by the difference between the two probabilities $(2p(\mu)-1)$. As clearly seen, the evolution of the growth rate takes place in a directed manner up to $\mu\approx 1$, whereas the dynamics of the evolution resembles the random walk when $\mu\gg 1$. This qualitative difference in the evolution dynamics above and below $\mu\approx 1$ may consistently describe the crossover of the survival time $T_s$, which is happening at around $\langle \mu_c\rangle \approx 1$. Namely, when the critical growth rate $\langle \mu_c\rangle$ is below one for long enough $\lambda$, the extinction happens relatively quickly since the growth rate systematically increase through the evolution, but when $\langle \mu_c\rangle$ is above one, the evolution takes a lot longer time due to the diffusive behavior, hence survival time $T_s$ grows faster as decreasing $\lambda$. 

Where does this transition from the directed to neutral evolution come from? The simple map dynamics (Eq.(\ref{eq:sol_map})) just tells us that the species with the largest $\mu/\gamma$ dominates the population which does not explain the random walk-like behavior of the averaged growth rate. In order to gain more insights, we studied the detailed dynamics of the take over events of the population by dominant species in stochastic simulation. 

Since the main purpose of this simulation was to ask how the dominant species changes one to another, we simulated the model with only two species which have a slightly different growth rate to each other.
One has a growth rate $\mu_l$ and the other has a higher growth rate, $\mu_h$ given as  
$\mu_h=1.05\cdot \mu_l$. Fig.\ref{fig:fig3}b shows the time-averaged populations of the species are plotted against the growth rate of the slowly-growing species ($\mu_l$). The fast-growing species dominates the whole population and the number of individuals is much larger than that of the slow grower in the small $\mu_l$ region, whereas the difference of the population sizes of the two species shrinks as $\mu_l$ increases and it gets indistinguishably small at $\mu_l\approx 1$. Examples of time courses are plotted in Fig.\ref{fig:fig3}c. The dynamics with large $\mu_l$ shows rather stochastic changes between the fast-grower dominating and slow-grower dominating states, while with small $\mu_l$ the fast grower is stably dominates the system.

This shrinkage of the gap in the two populations explains the transition from the directed to the neutral evolution of the growth rate.  Intuitively, the mechanism of this shrinkage can be understood by considering the effective fitness $\mu/\gamma$. Since we use the linear trade-off $\gamma(\mu)=a+b\mu$, 
the difference of the effective fitness between the fast species with the growth rate $\mu_h=(1+\delta)\mu_l$ and 
the slow species with the growth rate $\mu_l$ is given by $\mu_h/\gamma(\mu_h)-\mu_l/\gamma(\mu_l)=\delta a \mu_l/[(a+b\mu_l)(a+b(1+\delta )\mu_l]$, which approaches zero as $\mu_l$ increases. In other words, the larger the value of $\mu_l$ is, the harder it becomes for the fast species to take over the population. As a result, the growth rate performs almost a random walk through evolution for the large value of $\mu_l$. 

The more quantitative understanding can be obtained by applying the Wright-Fisher (WF) model \cite{ewens2012mathematical}. The WF model is a stochastic model describing temporal changes of the population structure such as the fixation probability and the fixation time. While the set-up of the present model does not fully fit the WF framework, we can apply the framework to the model with some assumptions which are described in the Appendix. The WF framework enables us to calculate the probability of the fast grower to be fixed in the population under no-mutation no-migration condition. The fixation probability is given by  
\begin{equation}
p_{\rm fix}=(1-\exp(-N_tuy))/(1-\exp(-N_tu)),\label{eq:fix_prob}
\end{equation}
where $N_t$, $y$ and $u$ represents the total number of the cells, the initial fraction of the fast growers, and the relative fitness of the fast grower defined as $u=(\mu_h/\gamma_h-\mu_l/\gamma_l)/(\mu_l/\gamma_l)$, respectively.

Fig.\ref{fig:fig3}d shows the comparison between the fixation probabilities computed by the simulation of the present model when the initial fraction of the fast-growing population $y$ set to 0.05 \footnote{To compute the fixation probability of the present model in the stochastic simulation, we set the mutation rate and the migration rate to zero, and run the dynamics from the fixed initial value of $N_l$, $N_h$, and $S$. A single run finishes if one of them extincts, and it is repeated with different random number seeds to compute the fixation probability.} and $p_{\rm fix}$ from the WF model in Eq.~(\ref{eq:fix_prob}). In the WF model, the total population size $N_t$ is replaced by the steady-state average population size of one species case for the given parameters, calculated from master equations with assuming long enough $\lambda$ so that the system typically reaches zero nutrient state in the famine period (Appendix D).    The two results show good correspondence. From the analytic expression of the fixation probability obtained from the WF approach (Eq.(\ref{eq:fix_prob})), we can see that the decrease of the fixation probability is led by two effects, namely, the decrease of the relative fitness advantage and the population size-effect. One effect is the form of $u$ being a decreasing function of $\mu$, hence as discussed before the advantage of the fast growth is reduced even the population size stays constant. In addition, the population size shrinks as the growth rate increases, and it makes the population dynamics noisier, making the small fitness difference no longer be the determinant of the dynamics. The two effects similarly contribute to the change of the fixation probability as shown by the dashed lines in Fig.\ref{fig:fig3}d, where the fixation probability Eq.~(\ref{eq:fix_prob}) with a constant relative fitness $u$ or a constant total population $N_t$ are also plotted. 

The WF model also in principle shows the parameter dependence of the crossover point, which should correspond to the point where fixation probability is sufficiently close to $1/2$ when starting from the equal population ($y=1/2$). 
The closed-form is difficult to obtain because the complex dependence of $N_t$ on $\mu$, but the form indicates that the crossover growth rate depends on the trade-off function parameter values ($a$ and $b$).

 \subsection*{Supplying more substrates leads to the quick extinction}
\begin{figure}[htbp]
\begin{center}
\includegraphics[width = 120 mm, angle = 0]{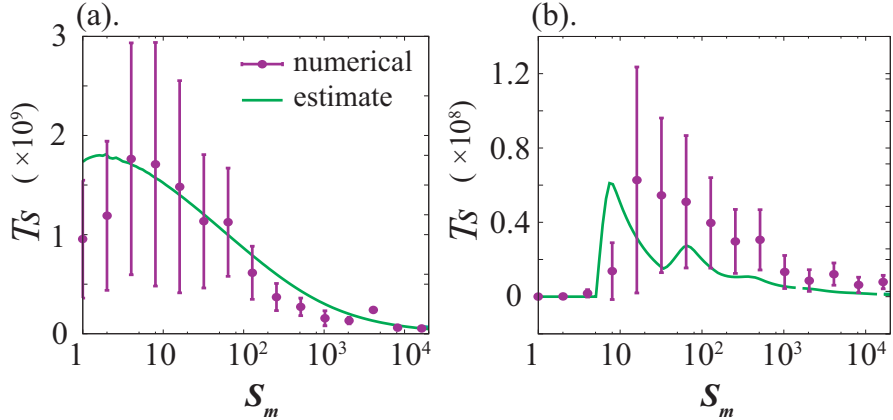}
\caption{The averaged survival time is plotted against $\Sm$ with a constant $\lambda$ value ($\lambda=10$ for (a), and $\lambda=100$ for (b)). While the survival time increases with $\Sm$, above a certain value of $\Sm$, the further increase of $\Sm$ causes the decrease of the survival time. The analytic estimate $\Sigma(1-p_{\rm fix})/p_{\rm fix}\cdot \tau_m$ is overlaid for each figure and captures non-monotonic behavior. Each point is obtained from $128$ independent evolution-extinction time courses and the error bars indicate the standard deviation. The parameters are set at $a=10^{-3},b=10^{-1},\rho=10^{-6}$, and $\delta \mu=10^{-2}$. $\epsilon$ is set at $0$ for the numerical simulation, while it is set at $10^{-12}$ for the analytic estimate because of the reason explained in the main text.}
    \label{fig:fig4}
  \end{center}
\end{figure}

 Finally, we show that an increase of the substrate supply $\Sm$ with fixed $\lambda$, hence increasing the average nutrient supply $\bar S$, decreases the survival time $T_s$. This is shown in Fig.~\ref{fig:fig4}, where the survival time $T_s$ is plotted as a function of $\Sm$ with constant $\lambda$. For the small $\Sm$ region, the survival time of the population increases as $\Sm$ gets larger because the amount of the substrate supply is too small at the left edge of the horizontal axis. On the other hand, the further increase of $\Sm$ shortens the survival time even though the waiting time is kept constant meaning that the total supply of the substrate increases.
 
 To ask what makes the survival time shorter, we estimated the survival time analytically. The survival time is approximated by the time needed to evolve the growth rate from the initial low value to the critical value with which the average population is close to one. We hypothesized that the survival time consists of the two main parts, namely, the time for gaining the new species by mutation $\tau_m$ and the time for the new species to take over the whole population.
 
  The bacterial population needs to wait that the fast grower appears by mutation to evolve. The mean time of the emergence of the grower by mutation is given as $\tau_m=1/(1-(1-\rho/2)^N)\approx2/\rho N$, where $N$ is the population size which depends on the population structure. It is possible that the slow grower appears and takes over the population, but we ignore this small possibility for simplicity. Then, after the fast grower appears, it either takes over the population or is eliminated. The time needed for the take over ($\tau_h$) and elimination of the mutant ($\tau_l$) are estimated as the fixation time in the WF framework. 
  
  The fast grower is expected to fail the fixation $(1-p_{\rm fix})/p_{\rm fix}$ times on average, and new mutant need to appear at every fixation failure. Therefore, the time for the dominant species to become $n$ to $n+1$ is given by
 \begin{equation}
 T_{n,n+1}=\frac{1-p_{\rm fix}}{p_{\rm fix}}(\tau_m +\tau_l)+\tau_h.\label{eq:Ts}
 \end{equation}
Note that $p_{\rm fix}$ , $\tau_h$, and $\tau_l$ are the functions of $\mu_n$ and $\mu_{n+1}$. In addition, at the every change of the dominant species that increase the average growth rate, the average population size decreases. This population size was already calculated in Appendix D from the master equation, and we assume that the extinction occurs when this population size becomes smaller than unity. Then,  by summing up $T_{n,n+1}$ over $n$, until the population size reaches to unity,  we obtain the estimate of the survival time.
 
 With the present parameter values shown in Fig.\ref{fig:fig4}, the dominant term of $\sum_n T_{n,n+1}$ is the time for the fast grower appearance by mutation, i.e., $\Sigma(1-p_{\rm fix})/p_{\rm fix}\cdot \tau_m$.   The comparison between the simulation and this expression is compared in Fig.~\ref{fig:fig4} shows a reasonable agreement. Also, the agreement indicates that the reduction of the survival time $T_s$ is mainly due to the increased rate of getting a fast-growing mutant with increasing $\Sm$ because it increases the typical population size. The more detailed comparison with the rest of the terms is given in Appendix F. \\
 


\section*{Discussion}
Here, we developed a stochastic population dynamics model in which the vital substrates were supplied to the environment by discrete, stochastic events rather than continuously. This stochastic substrate addition separates the dynamics into two phases, namely, the feast and the famine phase. During the feast period with plenty of substrates, the cells with a higher growth rate increase their population more quickly than the others. On the other hand, during the famine period, the cells could not grow but die due to the lack of substrates. 

With a trade-off between the growth rate and the growth yields the feast-famine cycle always led to the Tragedy of the Commons Dilemma-type result, while the outcome crucially depends on the form of the trade-off in the growth-death trade-off case.

The survival time, or the average time to extinction, was shown to have a power-law dependency to the average waiting time $\lambda$ with the constant time-averaged substrate supply $\bar{S}=\Sm/\lambda$, with a cross-over from close to $\lambda^{-2}$ to $\lambda^{-1}$. The cross over stemmed from the transition from the directed evolution to the undirected, random walk-like evolution dynamics. As the average growth rate increases, the difference in the fitness between two species with similar growth rates reduces. In addition, the population dynamics become noisier due to the decrease in the average population, and thus, the difference in fitness becomes less influential to the dynamics. Finally, it was shown that a pure increase of supplied substrate per nutrition addition event enhanced the extinction, and the reduction of the mutant appearance time due to increased population size was turned out to be the main part of the decrease of

For the cross-over from the directed to the random evolution to occur, the form of the trade-off is probably essential. With the square trade-off, one can choose the parameter values so that the evolution stops without enough shrinkage of the fitness gap and the population size, and then, only the directed evolution is expected. On the other hand, with the trade-off without the upper bound of the fitness, the population size becomes considerably small before the extinction. Also, the fitness gap always shrinks as the growth rate increases. By this argument, a cross-over being similar to what we presented above is expected to occur. 

\tcr{
It is worth noting that the trade-off between the growth rate and death rate (killing rate) are well reported under  a variety of stress conditions for bacteria \cite{zakrzewska2011genome,dong2009polymorphism,king2004regulatory,maharjan2013form,ferenci2016trade,porter2013trade,yang2019temporal,maharjan2013form,notley2002rpoS}, and the trade-offs between the growth and death rate are linear in some cases \cite{lee2018robust,tuomanen1986rate}. Also, the linear relationship between the growth- and the death rate is documented in the continuous culture of the fission yeast \cite{nakaoka2017aging}.  With the linear trade-off, the evolution could not find an optimal point in the present model and the whole population faced extinction. 
}

Undoubtedly, bacteria do not go extinct so easily as the present model has shown. Indeed, it is reported that the long-term evolution experiment of {\it E. coli} leads to the coexistence of the variety of bacterial genotypes \cite{lenski1991long,bouma1988evolution,good2017dynamics}.
Since the present model is a simple toy model, there is a large gap between the model and the real experiment. For the further investigations of bacterial population dynamics under the feast-famine cycle, it might be fruitful to argue the differences between them. 

First of all, the present model has no spatial degree of freedom. It is well-known fact that the introduction of the spatial structure will allow replicator models to have coexistence solutions \cite{mollison1977spatial,hufkens2009ecotones,mathiesen2011ecosystems,kerr2006local,schrag1996host,heilmann2010sustainability}. Introduction of the spacial structure was also proposed to avoid the TOC caused by the growth-yield trade-off \cite{pfeiffer2001cooperation}. Furthermore, it was experimentally shown that {\it E. coli} cells with the attenuated $rpoS$ outcompeted the wild-type cells in the well-mixed culture but coexisted with the wild-type in the spatially-structured culture \cite{hol2013spatial}. With the spatial structure introduced, it is possible that the present model shows the coexistence of multiple species that work to prevent extinction due to the TOC scenario.

Even without the effect of the space, it is possible that the physicochemical limits of the rates and the yield simply stop their values keep evolving. Another possible way to stop the extinction is to have the growth-death trade-off relationship being stronger than linear. But we would like to point out that having a stronger trade-off might be a weak strategy. Let us consider the situation where the trade-off relationship also evolves. In the population level, having the stronger trade-off relationship is preferential solution to avoid the extinction, but na\"{i}vely thinking, the bacterial population with a strong trade-off relationship is fragile to the invasion of another population with weaker trade-off relationship in the same sense as that slow growers are competed out by fast growers in the present model.

It is also worth noting that, in reality, a small fraction of the bacterial population can be non-growing persisters \cite{balaban2004bacterial,veening2008bistability,wakamoto2013dynamic}. The introduction of the persister phenotype makes the big jump between the high-growth and high-death and low-growth and low-death state possible, and it might change the evolutionary strategies. Also, the lag and stationary phases were not implemented in the model. The period of the lag phase is considered to increase with the starvation time (or the time in the stationary phase) \cite{merritt2018frequency,levin2010automated,himeoka2017theory}, and the prolonged lag time is clearly disadvantageous for the competition for substrates. The real bacteria might design their growth strategy also to cope with the length of lag time. Furthermore, it has been shown that during the death phase, the substrate provided by dead cells are utilized by alive cells to survive longer \cite{schink2019death}, and this feedback from cell death to the environment can be another factor to be considered to compare with the reality. 

Another factor to consider is the stochasticity in the fixation process. 
It is theoretically shown by using a Wright-Fisher-type model that the trade-off between the fitness advantage in WF scheme (e.g., the growth rate) and the carrying capacity (the maximum population of given species) makes a fixation probability of the species with smaller fitness advantage higher than that of the species with bigger fitness advantage under a certain condition \cite{houchmandzadeh2015fluctuation}. When a single cell of fast-growth and small carrying capacity is introduced into a community formed only of the other type, the chance of the former species to be selected is small due to a small frequency in the community. By contrast, in the opposite case, the population size of the community is small, and thus, the frequency and the fitness advantage are less influential to the selection. There, the randomness dominates the selection process rather than other factors. 

The effect of the feast-famine cycle to the bacterial evolution has been addressed experimentally, though typically the famine period is the order of a day or shorter that the bacteria enters the stationary phase but not the death phase. In Lenski and coworkers’ evolutionary adaptation experiment \cite{lenski1991long,wiser2013long,good2017dynamics,blount2012genomic,rozen2000long,blount2008historical} has been continued  for more than 30 years and a variety of mutations have been confirmed. For instance, some of the new genotypes can use citrate as a sole carbon source which the ancestral strain was unable to utilize \cite{blount2008historical,blount2012genomic} and cross-feeding polymorphism to appear \cite{rozen2000long}. Those observations mean that new niches were created during the evolution time series. As only a single nutrient is considered in the present model, there is only one niche and the possibility of the cross-feeding was not taken into account. An extension of the model to have more than two types of the nutrient and secretion of the chemicals from the cells will allow the model to have more niches and hence polymorphism to appear.

In a recent experiment of bacteria under repeated feast-famine cycle with the exchange of medium \cite{merritt2018frequency}, a large amount of the bacteria cells were also flushed out when the fresh media is added. In this case, the cells that form aggregates are selected, even though the growth was in the liquid culture. Though this particular experiment imposed the selection for aggregates, in general one should have in mind that the aggregation and hence spatial heterogeneity can arise even in a liquid culture \cite{kragh2018inoculation}, which can make a deviation from the prediction of a "well-mixed" model.     

Having these possible deviations from the present model in mind, it will still be interesting to perform an evolution experiment with longer famine period, to see the role of the trade-off between the growth rate and the death rate in the repeated feast-famine cycle.

As discussed, the present model has plenty of choices to be extended for emulating the strategy of the real bacterial population. But in spite of its simplicity, it provides several insights into how the feast-famine environment could affect the bacterial population dynamics, which hopefully helps the future development of our understandings of bacterial population dynamics and evolution.

\begin{acknowledgements}
The authors thank Nen Saito for fruitful discussions and Hiroshi Kori for suggesting the reference \cite{yang2019temporal}.  
This work was funded by the Danish National Research Foundation (BASP: DNRF120).
\end{acknowledgements}

\appendix
\section{Model equation for the multi-species model}
\begin{itemize}
\item $N_i$ : the number of the cell of $i$th phenotype
\item $S$ : the amount of the nutrient (max $\Sm$) 
\item $\Delta S_i=\lceil 1/Y_i\rceil$: the amount of the nutrient needed by the $i$th species to proliferate 
\item $\mu_i$ : growth rate of $i$th phenotype
\item $\gamma_i$ : death rate of $i$th phenotype
\item $\epsilon$ : migration (invasion) rate (normally it is set to zero)
\item $\lambda$ : the average waiting time for the substrate addition
 \end{itemize}
\begin{eqnarray}
\frac{dP({\vec N},S)}{dt}&=&\sum_{i}\mu_i\Bigl(\hat{\delta}_{S,\Sm}\hat{\delta}_{N_i,0}\E^{-1}_{N_i}\E^{\Delta S_i}_S(1-\rho+(\delta_{i,0}+\delta_{i,M-1})\rho/2)-(1-\theta(S))\Bigr) N_iP({\vec N},S)\nonumber\\
&+&\hat{\delta}_{S,\Sm}\frac{\rho}{2}\sum_i\E^{-1}_{N_i}(\E_S^{\Delta S_{i+1}}\hat{\delta}_{i,M-1}\mu_{i+1}N_{i+1}+\E_S^{\Delta S_{i-1}}\hat{\delta}_{i,0}\mu_{i-1}N_{i-1})P({\vec N},S)\nonumber\\
&+&\theta(S)\sum_{i}\gamma_i(\E_{N_i}-1)N_iP({\vec N},0)-P({\vec N},S)/\lambda+\delta_{S,\Sm}\sum_{s=-\infty}^{\Sm}P({\vec N},s)/\lambda \nonumber\\
&+&\epsilon(\E^{-1}_{N_i}-1)P({\vec N},S)\label{eq:meq_multi}
\end{eqnarray}
, where $\E_i$ is the step operator of species $i$ which acts to an arbitrary function $f(\cdots,N_i,\cdots)$ as $\E_i f(\cdots,N_i,\cdots)=f(\cdots,N_i+1,\cdots)$. Here, we allow to $P(N,S)$ with negative $S$ to have non-zero value because instead of setting the boundary at $S=0$ because it needs an exception handling, for instance, assuming that all the transition from such states go to the state with $S=0$. While the state $S<0$ is reachable, the right hand side of the equation is the same among $S\leq 0$. Therefore, by assuming the sum of the probabilities for $S\leq0$ represents the probability of the starving state, we can avoid the introduction of the exception handling. 

\section{Derivation of the map system}
Here, we derive the map system, first for the one-species case. To derive it, we ignore the interactions among species (i.e., deal with the growth/death dynamics of only one species.) and stochasticity in the bacterial growth/death dynamics. First, we calculate the duration of $\tau$ in which the bacterial population use up the newly added $\Sm$ substrates. Since the bacterial population grows exponentially as long as the substrate remains, the number of the bacteria at time $t$, $\tilde{N}(t)$ is given as 
$$\tilde{N}(t)=\tilde{N}(0)e^{\mu t},\ \ (0\leq t<\tau),$$
where we set $t=0$ as the time the substrates are newly added. The integral of this equation from $t=0$ to $t=\tau$ gives us the cumulative consumption of the substrate given by $(e^{\mu \tau}-1)\tilde{N}(0)/Y$. By the definition of $\tau$, the cumulative consumption equals to $\Sm$, and thus, we get 
\begin{equation}
    \tau=\mu^{-1}\ln(1+\Sm Y/\tilde{N}(0)), 
\end{equation}
Next, we introduce the waiting time between $n$th and $(n+1)$th substrate addition periods, $\Delta t(n)$. Note that if $\tau(n) < \Delta t(n)$, the bacteria run out all the substrates and start to die, otherwise, the population grows to $\tilde{N}(\tau)$. Thus, the population at the $(n+1)$th substrate addition event is determined by the population at the $n$th addition event as follows;
\begin{equation}
        N(n+1)=
    \begin{cases}
    N(n)e^{\mu \tau(n)}e^{-\gamma(\Delta t(n)-\tau(n))}\ (\tau(n)<\Delta t(n))\\
     N(n)e^{\mu \Delta t(n)},\\
    \end{cases}
    \end{equation}
where $N(n)$ indicates the number of the bacteria at the $n$th substrate addition. Since $\tau$ depends $N(n)$, we put $(n)$ to explicitly show that $\tau$ can have different values for different $n$'s. \\
Note that the average of $\ln(N(n+1)/N(n))$ over $\Delta t(n)$ with the exponential distribution gives the effective growth rate and leads to the deterministic time evolution of the bacterial population. The average results in
\begin{eqnarray*}
    N(n+1)&=&N(n)\exp(\hat{\mu}\lambda)\\
    \hat{\mu}(n)&=&\mu\Bigl(1-\exp(-\tau(n)/\lambda)\Bigr)-\gamma\exp(-\tau(n)/\lambda).
\end{eqnarray*}

For the multi-species case, the map system is derived in a similar way. It is given as 
\begin{eqnarray*}
    N_i(n+1)&=&N_i(n)\exp(\hat{\mu}_i(n)\lambda)\\
    \hat{\mu}_i(n)&=&\mu_i\Bigl(1-\exp(-\tau(n)/\lambda)\Bigr)-\gamma_i\exp(-\tau(n)/\lambda),
\end{eqnarray*}
where $\tau(n)$ is given as the solution of the following equation;
\begin{equation}
    \Sm=\sum_{i=0}^{M-1}N_i(n)/Y_i\Bigl(\exp(\mu_i\tau(n))-1\Bigr).
\end{equation}
Although we performed many numerical computation of the map dynamics with a variety of parameter choice, any attractor at which the species coexist was not found. Thus, we concentrate to the attractors at which only one species has non-zero population. By setting $\hat{\mu}_i(n)=0$ and $N_j(n)=0$ for any $ j\neq i$, we get the steady solution given by
\begin{equation}
    N_i^{\rm st}=\frac{\Sm Y_i}{(1+\gamma_i/\mu_i)^{\mu_i\lambda}-1}
\end{equation}
\section{The stability analysis of the map system}
In this section, we describe that the ratio between the growth rate and the death rate works as the fitness in the present model by performing the linear stability analysis of the map system which is derived in the previous section. \\
Here, we deal only with the fixed points at which only one species survives. Let us suppose that there are $M$ species. The Jacobi matrix of the system at the $n$th substrate addition event $(N_0(n),N_1(n)\cdots,N_{M-1}(n))$ is given as
 \begin{eqnarray}
 J_{ij}(n)&=&\frac{\partial N_i(n+1)}{\partial N_j(n)}=\delta_{ij}\exp[{\hat{\mu}_i(n)\lambda}]+N_i(n)\lambda\exp[{\hat{\mu}_i(n)\lambda}]\frac{\partial \hat{\mu}_i(n)}{\partial N_j(n)}.
 \end{eqnarray}
By renumbering the species index, we can assume that only the species with index $0$ survive at the attractor without losing generality. Since $N_0(n\to\infty)=N_0^{\rm st}$ and $N_1(n\to\infty)=\cdots N_{M-1}(n\to\infty)=0$ hold, the elements of the Jacobi matrix get simplified at the fixed point as
 \begin{eqnarray}
  J_{ij} = \begin{cases}
    \delta_{0j}+N_0^{\rm st}\lambda\frac{\partial \hat{\mu}_0}{\partial N_j}\Biggr|_{\rm st.} & (i=0) \\
    \delta_{ij}\exp[\hat{\mu}_i\lambda]\bigr|_{\rm st.}, & (i>0)
  \end{cases}
 \end{eqnarray}
 where $\cdot|_{\rm st}$ indicates the value of the functions at the steady state which we are interested in. This equation shows that the Jacobi matrix is triangle, and thus, the eigenvalues are 
 $$1+N_0^{\rm st}\lambda\frac{\partial \hat{\mu}_0}{\partial N_0}\Biggr|_{\rm st.},\ \exp[\hat{\mu}_1\lambda]\bigr|_{\rm st.},\ \exp[\hat{\mu}_2\lambda]\bigr|_{\rm st.},\cdots,\ \exp[\hat{\mu}_{M-1}\lambda]\bigr|_{\rm st.}$$ 
 The first eigenvalue $1+N_0^{\rm st}\lambda\frac{\partial \hat{\mu}_0}{\partial N_0}\bigr|_{\rm st.}$ is zero, and $\hat{\mu}_i|_{\rm st}$ is given as $\hat{\mu}_i|_{\rm st}=\mu_i(\gamma_0/\mu_0-\gamma_i/\mu_i)/(1+\gamma_0/\mu_0)$ because $\tau$ is given by $\lambda\ln(1+\gamma_0/\mu_0)$ holds at the fixed point. \\
 Since $\lambda$ is positive constant, $\gamma_0/\mu_0 < \gamma_i/\mu_i,$ for all $i=1,2,\cdots,M-1$ is the stability condition for this fixed point. Therefore, the only one fixed point with the largest $\mu/\gamma$ is stable among all the fixed points at which only one dominant species exists.        
\section{The approximated solution of the master equation}
\begin{figure*}[htbp]
\includegraphics[width = 120 mm, angle = 0]{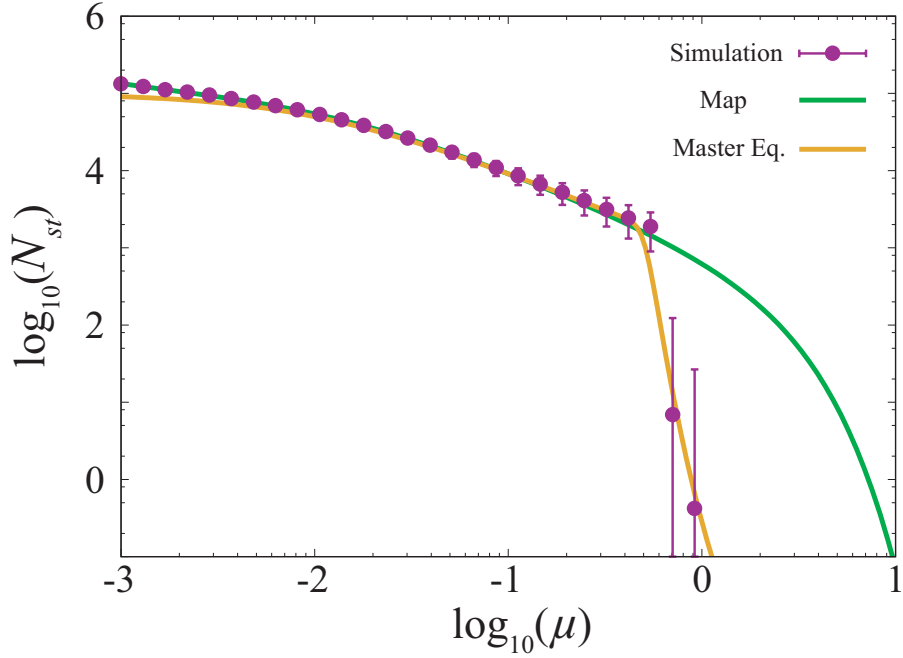}
\begin{center}
\caption{The comparison among three types of the analytic estimates of the steady state solution and the numerically computed steady state solution. "Map" and "Master Eq." corresponds to the steady state solution given in Eq.(\ref{eq:sol_map}) and Eq.(\ref{eq:sol_meq}), respectively. The "Map" solution fails to correctly predict the sudden drop of the steady population due to the deterministic and continuous approximation  of the variables and averaging of the waiting time, while the "Master Eq." cannot capture the steady solution correctly in low growth rate region because we assumed the growth rate is sufficiently larger than $1/\lambda$ to calculate the master equation. Parameter values are set to $\lambda=10,\Sm=10^3,\ep=10^{-8},a=0.1$ and $b=10^{-3}$}
    \label{fig:App_sol}
  \end{center}
\end{figure*}

In this section, we calculate the steady state solution of the master equation with only one species Eq.(\ref{eq:single}). Here, we write the probability of the each state as $P_S(N)$ instead of $P(N,S)$ and convert the single master equation into the system of $(1+\Sm)$ master equations just for the readability of following calculations. For the sake of simplicity, here we consider only the cases which satisfy $\Sm \lceil Y\rceil \in {\mathbf Z}$. Then, the difference of the yield is absorbed into the value of $\Sm$ which allows us to set the yield to unity. \\
The full-model is approximated by assuming that $\lambda\gg\ln(1+\Sm/N_{\rm st})/\mu$ holds to ignore the terms $P_S(N)/\lambda$ for $S>0$. Here, we also introduced a spontaneous migration with rate $\epsilon$, because the master equation has an absorbing state at $(N,S)=(0,\Sm)$ and without the migration term, only the steady state is $P(0,\Sm)=1$ and $0$ for the others. The approximated master equations are given as  
\begin{eqnarray*}
\frac{dP_\Sm(N)}{dt}&=&-\mu NP_\Sm(N) + \epsilon \Bigl(P_\Sm(N-1)-P_\Sm(N)\Bigr)+P_0(N)/\lambda\\
\frac{dP_S(N)}{dt}&=&\mu(N-1)P_{S+1}(N-1)-\mu NP_S(N) + \epsilon \Bigl(P_S(N-1)-P_S(N)\Bigr),(S=1,\cdots \Sm-1)\\
\frac{dP_0(N)}{dt}&=&\mu (N-1)P_1(N-1) + \epsilon \Bigl(P_0(N-1)-P_0(N)\Bigr)\\&+&\gamma \Bigl((N+1)P_0(N+1)-nP_0(N)\Bigr)-P_0(N)/\lambda
\end{eqnarray*}
Here, we introduce the moment-generating function for each $S$ defined as $$G_S(z)=\sum_{N=0}^{\infty}z^NP_S(N).$$ We get following equations at the steady state
\begin{eqnarray}
0&=&-\mu zG'_\Sm+\ep G_\Sm(z-1)+G_0/\lambda \nonumber \\
0&=&-\mu z G'_S+\mu z^2 G'_{S+1}+\ep G_S(z-1),\ (S=1,\cdots,\Sm-1) \label{eq:moment0}\\
0&=&\mu z^2G'_1-\gamma(z-1)G'_0+\ep G_0(z-1)-G_0/\lambda\nonumber,
\end{eqnarray}
where $G'_S$ represents the first-order derivative of $G_S$'s respect to $z$. By setting $z=1$, we get
\begin{eqnarray}
0&=&-\mu zG'_\Sm +G_0/\lambda  \nonumber\\
0&=&-\mu z G'_S+\mu z^2 G'_{S+1}\ (S=1,\cdots,\Sm-1)  \label{eq:moment0_z1}\\
0&=&\mu z^2G'_1-G_0/\lambda. \nonumber
\end{eqnarray}
Thus an equality 
\begin{equation}
G'_1(1)=G'_2(1)=\cdots=G'_\Sm(1)=G_0(1)/\mu\lambda \label{eq:result0}
\end{equation}
holds.\\
Also, by differentiating Eq.(\ref{eq:moment0}) respect to $z$ again, we get
\begin{eqnarray}
0&=&-\mu G'_\Sm -\mu z G''_\Sm+\ep G_\Sm+\ep G_m(z-1)+G'_0/\lambda\nonumber\\
0&=&-\mu G'_S-\mu z G''_S+2\mu z G'_{S+1}+\mu z^2 G''_{S+1}\nonumber\\&+&\ep G_S+\ep G'_S(z-1)\label{eq:moment1}\\
0&=&2\mu zG'_1+\mu z^2G''_1-\gamma G'_0-\gamma(z-1)G''_0\nonumber\\&+&\ep G_0+\ep G'_0(z-1)-G'_0/\lambda.\nonumber
\end{eqnarray}
By summing up these $\Sm$ equations and setting $z=1$, we get
\begin{eqnarray}
0&=&\ep + \mu\sum_{S=1}^\Sm G'_S-\gamma G'_0 \nonumber\\
\Leftrightarrow 0&=&\ep +M\frac{G_0}{\lambda}-\gamma G'_0 \label{eq:result1}
\end{eqnarray}
From Eq.(\ref{eq:result0}) and (\ref{eq:result1}), the first moment is given by 
\begin{eqnarray}
\sum_{S=0}^\Sm G'_S(z=1)&=&G'_0(1)+\sum_{S=1}^\Sm G'_S(1)\nonumber\\
&=&\frac{\ep}{\gamma}+M\frac{G_0(1)}{\lambda\gamma}+M\frac{G_0(1)}{\lambda\mu}\nonumber\\
&=&\frac{\ep}{\gamma}+\frac{MG_0(1)}{\lambda}\Bigl(\mu^{-1}+\gamma^{-1}\Bigr)\label{eq:reault2}
\end{eqnarray}
Since $G_0(1)$ is indeterministic from the equations (\ref{eq:moment0_z1})-(\ref{eq:result1}), we make an anzats that $G_0(1)$ has the form 
\begin{equation}
G_0(1)=\frac{\lambda}{\lambda+H_0(\Sm)/\mu+\ep^{-1}\exp(-H_1(\Sm)/\gamma\lambda)}, \label{eq:G0}
\end{equation}
with 
\begin{eqnarray*}
H_0(\Sm)&=&\sum_{n=\lceil\hat{N}(\Sm)\rceil}^{\lceil\hat{N}(\Sm)\rceil+\Sm-1}\frac{1}{n}\\
H_1(\Sm)&=&\sum_{n=1}^{\lceil\hat{N}(\Sm)\rceil+\Sm}\frac{1}{n}.\\
\end{eqnarray*}
Here, we use the steady state solution the map dynamics in the main text Eq.(\ref{eq:sol_map}) as $\hat{N}$ and $\lceil\cdot\rceil$ represents the ceiling function.\\
At the end, the steady state solution of the bacterial population is given by
\begin{eqnarray}
N_{\rm st}=\sum_{S=0}^{\Sm}G'_S(1)=\frac{\epsilon}{\gamma}+\Sm\frac{1/\mu+1/\gamma}{\lambda+H_0(\Sm)/\mu+\exp[-H_1(\Sm)/\gamma\lambda]/\ep}.\label{eq:sol_meq}
\end{eqnarray}
Fig.\ref{fig:App_sol} shows the comparison of the two types of analytic estimates of the steady state solution to the numerically computed one (time-averaged population). The solution derived in this section gets less accurate when $\mu$ is small due to the assumption $\lambda \ll \mu N_{\rm st}$. On the other hand, it predicts the extinction of the population (the strong drop of the population) precisely.  \\\\
We made the anzats that $G_0(1)$ has the form shown in Eq.(\ref{eq:G0}) because $G_0(1)$ represents the fraction of time that the system stays at the states with no substrate. Apparently, the timescale to escape from the no-substrate states is $\lambda$. On the other hand, the length of time that the system stays at the with-substrate states is given by $H_0/\mu$. Suppose that the system is at the state with $(N,S)=(n,\Sm)$. The spontaneous immigration rate is ignorable if there are the substrates, and thus, the bacteria proliferates without experiencing any other elemental processes until all the substrates are run out. Since the number of bacteria increases strictly one by one as a single substrate is used, the timescale for reaching to the non-substrate states is obtained by summing up the timescales of the proliferation with given number of the substrate. It is given as $(1/n+1/(n+1)+\cdots+1/(n+\Sm-1))/\mu$. The number of the bacteria right after the substrate addition is typically given by $\lceil \hat{N}\rceil$, and thus, the sum is written as $H_0/\mu$.\\
Lastly, there is finite possibility that the system reaches the state $(N,S)=(0,\Sm)$ which becomes the absorbing state under $\epsilon\to0$ limit. Once the system falls into the state, the only one elemental process could take place is the spontaneous immigration occurring at rate $\epsilon$. It might be natural to estimate the probability of the system to fall into this state from the ratio of two timescales, namely, the timescale of the substrate addition and cell death. A single bacterium in $N$ bacteria dies within $1/\gamma N$ on average. Thus, the whole population is expected to extinct within $H_1/\gamma$. Therefore, the probability that the total population becomes zero before the  substrates being supplied can be approximated by $(\exp[-H_1/\gamma\lambda])$.\\
The sum of these three parts gives the typical length of time consumed in between two substrate addition events, and thus, the ratio between $\lambda$ and the sum corresponds to what we desired.\\ 

\section{The fixation probability and the fixation time calculated from the Wright-Fisher model}
In this section, we explain how the WF model is applied to the present model. Before listing up the assumptions made, we briefly explain the framework of the WF model, while detailed descriptions can be found in standard textbooks \cite{ewens2012mathematical}. The WF model is a Markov-chain model which predicts the temporal evolution of relative abundances of the species. Suppose that there are two species, species $h$ and $l$ (high and low), which has the relative fitness $1+u$ (here we assume $u>0$) and $1$, respectively. The total population is kept to the constant number $N_t$, and the generational change is assumed to take place simultaneously. The probability that one individual at the next generation to be the species $h$ is given by $p_i=i(1+u)/[i(1+u)+(N_t-i)]$, where $i$ is the number of the species $h$ at the previous generation. Thus, the transition probability

\begin{center}
From (the number of h cells,the number of l cells) $=$ $(j,N_t-j)$ To $(i,N_t-i)$
\end{center}
in one generation is given by
 \begin{equation}
P_{ij}=\binom{N_t}{i}p^i_j(1-p_j)^i\label{eq:fixation_prob}.
 \end{equation}
If the mutation is not included in the model, the model has two absorbing states at which only one species dominates the whole population corresponding to the fixation of the species. By using the diffusion approximation of the WF model, the fixation probability of species $h$ is analytically given as $p_{\rm fix}=(1-\exp(-N_tuy))/(1-\exp(-N_tu))$, where $y$ is the initial abundance of the species $h$ relative to the total population.\\
For fitting the present model into the framework of the Wright-Fisher (WF) model, we have to make several assumptions, while they are not always fulfilled by the present model exactly. Here, we first list the assumptions which we need to make with comments.\\
{\bf (i). The total number of the population is constant} : Clearly this requirement is violated, while we assume that the total number of the population is fixed at the steady-state value (Eq.(\ref{eq:sol_meq})) with the growth rate of the slow grower. \\
{\bf (ii). Proliferation/Cell death takes place simultaneously} : Since cell division and cell death occur as a Poisson random event in the present model. We can neither fulfill the two conditions. We assume, however, that it is effectively satisfied by coarse-graining the time. We suppose that on average every $N\lambda/\Sm$ time interval, $N$ individuals die and $N$ bacteria are newly born, and accordingly, one generation in the WT framework corresponds to $N\lambda/\Sm$ in the timescale of the present model.\\   
{\bf (iii).The fitness function} : Due to the differences between the WF framework and the present model, it is unclear which parameter in the present model should correspond to the relative fitness advantage. So, here we adopt the ratio between the growth rate and death rate as the fitness function because it is already shown to determine the stable fixed point under the deterministic approximation of the model.\\\\
Here, we calculate the fixation probability and the fixation time of the two-species WF model. We assume that there are two species with slightly different growth rates and death rates. We do not include the mutation process in the analysis. Instead, we set a very small portion of the total population ($y$) as the bacteria with higher fitness. The fixation probability of the bacteria with the higher fitness is given as
\begin{eqnarray}
p_{\rm fix}(u,N,y)=\frac{1-\exp[-\alpha y]}{1-\exp[-\alpha]},
\end{eqnarray}
under the continuous (large $N$) limit, where $\alpha$ is defined as $\alpha=uN$. Since one of the two species is eventually fixed, the fixation probability of the other species is given as $1-p_{\rm fix}$. \\\\
Next, the fixation time $\tau_i\ (i=l,h)$ is given by solving the backward Kolmogorov equation below,
\begin{eqnarray*}
    &&\alpha u(1-u)\frac{\partial \tilde{\tau}_i}{\partial y}+\frac{1}{2N}u(1-u)\frac{\partial^2 \tilde{\tau}_i}{\partial y^2}=
    \begin{cases}
    -p_{\rm fix} & (i=h)\\
    -(1-p_{\rm fix}) & (i=l)\\
    \end{cases}\\
    &&\tau_h=\tilde{\tau}_h/p_{\rm fix},\ \tau_l=\tilde{\tau}_l/(1-p_{\rm fix}).
\end{eqnarray*}
The solutions are given by the following equations 
\begin{eqnarray}
\tilde{\tau}_l(u,N,y)&=&C_0(u,N)-C_1(u,N)e^{-2\alpha y}+F(u,N,y)+G(u,N,y),\\
\tilde{\tau}_h(u,N,y)&=&D_0(u,N)-D_1(u,N)e^{-2\alpha y}-F(u,N,y),\\\\
F(u,N,y)&=&\frac{1}{u(1-e^{-\alpha})}\Biggl[\ln\Bigl(\frac{y}{1-y}\Bigr)-\cE\bigl(-2\alpha y\bigr)+\cE\bigl(2\alpha(y-1)\bigr)\\&+&e^{-\alpha y}\bigl[\cE\bigl(-\alpha y\bigr)-\cE\bigl(\alpha y\bigr)\bigr]+e^{-\alpha y}\bigl[\cE\bigl(\alpha (y-1)\bigr)-\cE\bigl(-\alpha (y-1)\bigr)\bigr]\Biggr],\\
u\cdot G(u,N,y)&=&\ln\Bigl(\frac{1-y}{y}\Bigr)+\cE(-2\alpha y)-\cE(2\alpha(y-1)),\\
D_1(u,N)&=&(F(u,N,1)-F(u,N,0))/({1-e^{-2\alpha}}),\\
D_0(u,N)&=&D_1(u,N)+F(u,N,0),\\
C_1(u,N)+D_1(u,N)&=&(G(u,N,0)-G(u,N,1))/(1-e^{-2\alpha}),\\
C_0(u,N)+D_0(u,N)&=&(C_1(u,N)+D_1(u,N))-G(u,N,0),
\end{eqnarray}
where $\cE(x)$ is the product of the exponential function $e^x$ and the exponential integral function with the negative argument ${\rm Ei}(-x)$, i.e $\cE(x)=e^x\cdot {\rm Ei}(-x)=e^x\int_x^{\infty}e^{-t}t^{-1}dt$. 

\section{An analytic estimate of the survival time before extinction}
In this section, we derive the analytic estimate of the survival time before the extinction shown in Eq(\ref{eq:Ts}) in the main text. \\
With the sufficiently small mutation rate $\rho$, the survival time can be estimated from the recursive taking-over process between two species with different fitness values because there are only two neighboring species in the system most of the time. So, we calculate the average length of time needed for the $(i+1)$th species dominates the whole population which was initially dominated by the $i$th species. We denote this as $T_{i,i+1}$, and consecutively sum up them to get the total length of time.\\
The time for the $(i+1)$th species to take over the population consists of three parts, namely, time for the mutant appearance ($\tau_m$), time consumed by failures of the fixation ($i$th species dominates again, being given by $\tau_l$), and the time for the fixation ($(i+1)$th species dominates, being given by $\tau_h$). For making it clear that each part depends on the index of the species, we put index explicitly as $\tau_m^i, \tau_l^{i,i+1}$ and $\tau_h^{i,i+1}$, while omitting the superscript when we mention general features. In this section, we suppose the growth rate $\mu$ and the maximum number of the substrate $\Sm$ as parameters, and only $\mu$ and $\Sm$ are indicated as arguments of functions.\\
A single mutant with the higher fitness appears within $\tau_m=(1-(1-\rho/2)^N)\approx 2/\rho N$ generation on average. Note that we can ignore the appearance of the mutants with lower fitness because they typically fail to increase their portion in the whole population due to the low fitness and the small population. Thus, the result does not alter even if the $(i-1)$th species is dealt as the $i$th species. In the following argument, we assume that the initial portion of the bacteria with higher fitness, $y$, equals to the inverse of the population size, $N$, because it is less likely to happen that more than two mutant appears at the same time due to low mutation rate.\\
The expected number of failures of the fixation (the count of the event that a mutant appeared but eventually eliminated) is given by
\begin{eqnarray}
R(u(\mu_l,\mu_h),N(\mu_l,\Sm))&=&R(\mu_l,\mu_h,\Sm)\\
&\equiv& \Pi_l(u,N,1/N)/\Pi_h(u,N,1/N)\\
&=&\Bigl[\coth(u/2)\sinh(uN/2)-\cosh(uN/2)\Bigr]e^{-uN/2}
\end{eqnarray} 
At every failure, the population has to wait for a next mutant to appear and one failure takes time with length $\tau^{i,i+1}_l$. The fixation of the species with higher fitness has to take place only once, and thus, the population is expected to evolve their dominant species from $i$th to $(i+1)$th species in 
$$T_{i,i+1}(\mu_i,\mu_{i+1},\Sm)=R(\mu_i,\mu_{i+1},\Sm)\cdot\Bigl[\tau_m^i(\mu_i,\Sm)+\tau_l^{i,i+1}(\mu_i,\mu_{i+1},\Sm)\Bigr]+\tau_h^{i,i+1}(\mu_i,\mu_{i+1},\Sm).$$
Lastly, we need to convert the timescale because the unit of time is the number of generations in the WT model. Since $\Sm$ bacteria divide after a single substrate addition event, $N/\Sm$ substrate addition events are needed for a single generation to pass ($N$ bacteria divide). The substrate addition events take place every $\lambda$ interval on average. Thus, the timescale is converted from WT to the present model by multiplying $N(\mu)\lambda/\Sm$. \\
 
$$T_s(\Sm)=\frac{\lambda}{\Sm}\sum_{i=0}^{i^*}N(\mu_i)T_{i,i+1}.$$
Note that here we assumed that the $(i+1)$th species always has a fitness value higher than that of the $i$th species ($u>0$). This assumption makes it unlikely to happen that the $(i-1)$th species appear by mutation and eliminate the $i$th species. This directedness is vital to make an estimate of the survival time by simply summing up the length of time for each taking-over event $T_{i,i+1}$. The assumption is violated if the fitness function $\mu/\gamma(\mu)$ has a peak, for example, the fitness function with the square trade-off.

\begin{figure}[htbp]
\begin{center}
\includegraphics[width = 120 mm, angle = 0]{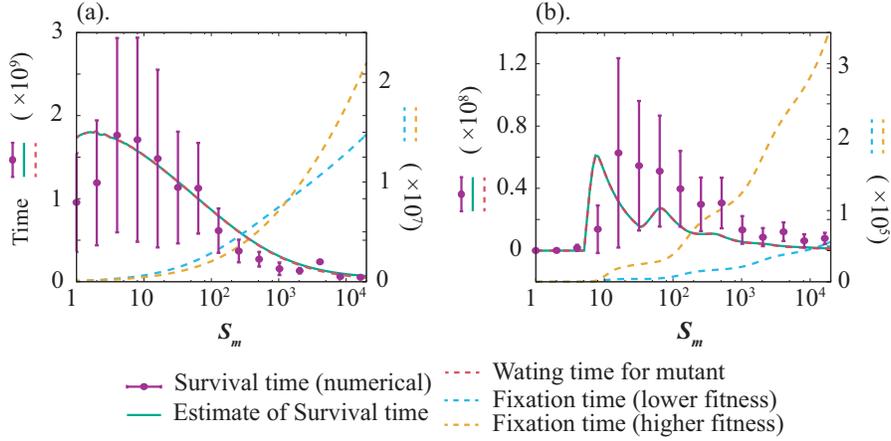}
\caption{The comparison of the average survival time $T_s$ and the estimate of the value. In this figure, the estimate is given by $\sum \Bigl[(1-p_{\rm fix})/p_{\rm fix}(\tau_m+\tau_l)+\tau_h\Bigr]$. Each part $\sum (1-p_{\rm fix})/p_{\rm fix}\tau_m$, $\sum (1-p_{\rm fix})/p_{\rm fix}\tau_l$, and $\sum \tau_h$ is plotted separately as red, cyan, and orange dashed line, respectively.  The parameters are set at $a=10^{-3},b=10^{-1},\rho=10^{-6}$, and $\delta \mu=10^{-2}$. $\epsilon$ is set at $0$ for the numerical simulation, while it is set at $10^{-12}$ for the analytic estimate because of the reason explained in the main text.}
    \label{fig:App_Ts}
  \end{center}
\end{figure}

\end{document}